\documentclass[conference]{IEEEtran}
\usepackage{amsmath,amsthm}
\usepackage{subfigure}
\usepackage{cite}
\usepackage{graphicx}
\usepackage{epstopdf}
\usepackage{color}
\usepackage{amsfonts,amsmath,amssymb}
\usepackage{cite}
\usepackage{graphicx}
\usepackage{url}
\usepackage{bm}
\usepackage{bbm}
\usepackage{amssymb}
\usepackage{braket}
\usepackage{soul}
\usepackage[T1]{fontenc}
\usepackage[colorlinks,linkcolor=blue,anchorcolor=blue,citecolor=blue]{hyperref}

\usepackage{fancyhdr}
\begin{document}	
\title{\huge{Multi-mode CV-QKD with \\ Noiseless Attenuation and Amplification}}
\author{
	\IEEEauthorblockN{Mingjian He$^1$,  Robert Malaney$^1$, and Benjamin Burnett$^2$}\\		
	\IEEEauthorblockA{$^1$School of Electrical Engineering  \& Telecommunications,\\
		The University of New South Wales,
		Sydney, NSW 2052, Australia. \\
		$^2$Northrop Grumman Corporation, San Diego,
		California, USA. }}

\maketitle
 \thispagestyle{fancy}
\renewcommand{\headrulewidth}{0pt}
	
\begin{abstract}
	
In this work we study the use of noiseless attenuation and noiseless amplification, in terms of multi-mode Continuous-Variable (CV) Quantum Key Distribution (QKD) over satellite-to-ground channels.
We propose an improved multi-mode CV-QKD protocol where noiseless attenuation and noiseless amplification operations are applied at the transmitter and the receiver, respectively.
Our results show that consistent with single-mode CV-QKD, the noiseless amplification operation, when applied at the receiver, can increase the transmission distance and the channel noise tolerance of multi-mode CV-QKD.
Different from single-mode CV-QKD, in multi-mode CV-QKD the key rate improvement offered by noiseless amplification can be further enhanced by adding noiseless attenuation at the transmitter.

\end{abstract}

\section{Introduction}

Increasing the key rate and the channel-loss tolerance of
Continuous-Variable (CV) Quantum Key Distribution (QKD)
is a topic of much ongoing research. One method to reduce the impact of channel losses is to perform noiseless attenuation at the transmitter. In this context, recent work on entanglement-based CV-QKD protocols with noiseless attenuation is of particular interest \cite{guo2019continuous,ye2019improvement,guo2020quantum}. 
In these works noiseless attenuation is realized by a zero photon catalysis operation, with the results showing that the attenuation can increase both the QKD key rate and the transmission distance\footnote{In \cite{guo2019continuous,ye2019improvement,guo2020quantum} noiseless attenuation is shown to increase the QKD key rate (transmission distance) only when some parameter of the initial entangled state (e.g., the squeezing) is not optimized for maximum key rate (transmission distance).}.

An alternative method to combat channel losses is to perform noiseless amplification at the receiver. 
Indeed, entanglement-based CVQKD protocols with noiseless amplification have been widely studied in recent years, demonstrating how the amplification can also increase the QKD key rate as well as the transmission distance \cite{gisin2010proposal,blandino2012improving,fiuravsek2012gaussian,walk2013security,wang2014improving,zhang2015noiseless,9024548,ghalaii2020discrete}).

However, all of the aforementioned studies on CV-QKD are under the assumption that each beam of the Einstein-Podolsky-Rosen (EPR) state only contains a single frequency mode.
In reality, any quantum state contains multiple frequency modes - an issue of increased concern when broadband pulses of light (ultra-fast pulses in the time domain) are utilized.
Multi-mode entangled states potentially allow for a higher quantum channel capacity \cite{christ2012exponentially,usenko2014entanglement}.
It is therefore natural to investigate what impact  noiseless attenuation and noiseless amplification can have on a CV-QKD system utilizing multi-mode entangled resources.

In our previous study we have investigated the performance of a multi-mode CV-QKD protocol using parametric down-converted (PDC) states with non-Gaussian operations over fixed-attenuation channels, determining which operation is preferred at the transmitter \cite{he2019multi}.
We have shown that in the multi-mode setting, non-Gaussian operations can improve the maximized key rate but cannot increase the maximal transmission distance.
{In this work we extend our previous study by introducing multi-mode noiseless attenuation and amplification.
We will focus on satellite-to-ground channels for multi-mode CV-QKD.
Recent advances in the satellite-based deployment of QKD \cite{liao2017satellite} represent a significant step forward in the creation of global-scale quantum networks.
However, it is important to further study QKD in this context searching for improvement in the communications set up.
For a review of CV quantum communications via
satellite see \cite{hosseinidehaj2018satellite}.}

Our contributions in this work are as follows.
(i) Noiseless attenuation and noiseless amplification in the multi-mode setting are investigated. 
(ii) We introduce an improved multi-mode CV-QKD protocol combining noiseless attenuation and noiseless  amplification.
(iii) We calculate the maximized key rate of the improved  protocol over satellite-to-ground channels, showing that the loss tolerance can be significantly enhanced.

The remainder of this paper is organized as follows. 
In Section~\ref{sec:amp} we discuss noiseless attenuation  and noiseless amplification in the multi-mode setting.
In Section~\ref{sec:qkd} we detail our improved multi-mode CV-QKD protocol.
Section~\ref{sec:channel} illustrates the satellite-to-ground channel model.
In Section~\ref{sec:results} we present and discuss our simulation results.

\section{Multi-mode noiseless attenuation and noiseless amplification}\label{sec:amp}
{ Consider an arbitrary multi-mode (broadband frequency mode) state expressed in the Fock basis},
\begin{equation}
\ket{\psi}=\sum_{n=0}^{\infty}a_n \ket{n},
\end{equation}
where 
\begin{equation}
\ket{n}=\frac{\hat{A}^{\dagger n}}{\sqrt{n!}}\ket{0},
\end{equation}
and $a_n$ is a normalized coefficient.
The creation operator on a broadband frequency mode is defined as
\begin{equation}
\hat{A}^\dagger=\sum_{m=1}^{\infty}\gamma_m a_m^\dagger,
\end{equation}
where $a_m^\dagger$ is the creation operator on a specific single-frequency mode (indexed by $m\in\left\{1,2,...,\infty\right\}$) and $\gamma_m$ is the  weighting coefficient.
Noiseless attenuation acting on $\ket{\psi}$ can be represented by a transformation

\begin{equation}\label{eq:ampeq}
\ket{\psi}\rightarrow  \sqrt{T}^{\hat{N}}\ket{\psi},
\end{equation}
where $T<1$ is the transmissivity of the noiseless attenuation operation and
$\hat{N}=\hat{A}^\dagger \hat{A}$.

In the multi-mode setting, noiseless attenuation can also be expressed by a transformation
\begin{equation}
\ket{\psi}\rightarrow  \sqrt{T}^{\tilde{N}}\ket{\psi},
\end{equation}
where 
\begin{equation}
\tilde{N}=\sum_{m=1}^{\infty} a_m^\dagger a_m,
\end{equation}
since 
\begin{equation}
\sqrt{T}^{\tilde{N}}\ket{n}=\sqrt{T}^{\hat{N}}\ket{n}=\sqrt{T}^n\ket{n}.
\end{equation}

Likewise, noiseless amplification can be represented by a transformation
\begin{equation}\label{eq:att_transform}
\ket{\psi}\rightarrow  G^{\hat{N}}\ket{\psi},
\end{equation}
where 
$G>1$ is the gain of the noiseless amplification operation.

\subsection{The application of noiseless attenuation and noiseless amplification to multi-mode Gaussian states}
A multi-mode Gaussian state may contain multiple orthogonal broadband frequency modes, each mode named a \textit{supermode}.
The parametric down-conversion (PDC) process is commonly used to create entangled states.
In reality, this process does not produce a single EPR state with single frequency modes, but an ensemble of independent EPR states with broadband frequency modes.
In the PDC process, a pump laser is first fed into a non-linear crystal. Two correlated beams, labeled $\bm{A}$ and $\bm{B}$, are then created.
Let $\hat{A}^\dagger_k$ and $\hat{B}^\dagger_k$ be the creation operators of the supermodes in beams $\bm{A}$ and $\bm{B}$, respectively, where the subscript $k\in\left\{1,2,...,\infty\right\}$ is used to index the supermodes. The output state of the PDC process can be written as \cite{christ2012exponentially}
\begin{equation}\label{eq:PDCinSch}
\begin{aligned}
\ket{\text{PDC}}_{AB} &= \bigotimes_{k=1}^\infty \exp\left[g \lambda_{k}\left(\hat{A}^\dagger_k \hat{B}^\dagger_k  - \hat{A}_k \hat{B}_k\right) \right]\ket{0}\\
&=\bigotimes_{k=1}^\infty \ket{\text{EPR}_k}_{AB},
\end{aligned}
\end{equation}
where
\begin{equation}\label{eq:EPRsingle}
\ket{\text{EPR}_k}_{AB}=\left(\sqrt{1-\tanh^2{r_k}}\right)\sum_{n=0}^\infty \tanh^n{r_k} \ket{n,n}_{AB},
\end{equation}
$r_k=g \lambda_{k}$ is the squeezing parameter, $g$ is the overall gain of the PDC process, and the $\lambda_{k}$'s are normalized coefficients, which follow an exponentially decaying distribution for the most likely PDC sources \cite{christ2012exponentially}.

The quadrature operators associated with one PDC state are defined as ($\hbar = 2$ is adopted)
\begin{align}\label{eq:quadopt}
\begin{array}{l}{\hat{X}_{k}^{A}=\hat{A}_{k}+\hat{A}_{k}^{\dagger}, \hat{P}_{k}^{A}=i\left(\hat{A}_{k}^{\dagger}-\hat{A}_{k}\right),} \\ {\hat{X}_{k}^{B}=\hat{B}_{k}+\hat{B}_{k}^{\dagger}, \hat{P}_{k}^{B}=i\left(\hat{B}_{k}^{\dagger}-\hat{B}_{k}\right).}
\end{array}
\end{align}
Being an ensemble of EPR states, the PDC state can be fully characterized by the covariance matrix (CM) of the quadrature operators in Eq. (\ref{eq:quadopt}). The CM of each EPR state has the form
\begin{align}\label{eq:CMepr}
\Sigma_{k}=\left(\begin{array}{ll}
{\cosh \left(2 r_{k}\right) I_2} & {\sinh \left(2 r_{k}\right) Z} \\ 
{\sinh \left(2 r_{k}\right) Z} & {\cosh \left(2 r_{k}\right) I_2}
\end{array}\right),
\end{align}
where $I_2$ is the 2-by-2 identity matrix and $Z=\text{diag}[1,-1]$. 

Since noiseless attenuation  and  amplification are both Gaussian operations \cite{fiuravsek2012gaussian},
to find the resultant state after the operations we only need to consider the evolution of the CM of the state.
{ Suppose a noiseless attenuation operation is applied to the first supermode ($k=1$) in beam $\bm{A}$ of a PDC state.}
{For clarity, we assume this state only has some finite number, $K$, of equivalent EPR states.
Let $\Sigma=\bigoplus_{k=1}^{K}\Sigma_k$ be the CM of the PDC state,}
an efficient way to derive the CM after noiseless attenuation is to employ the $Q$-function of the PDC state \cite{gagatsos2014heralded} 
\begin{align}
Q(r)=\frac{\sqrt{\operatorname{det}(\sigma)}}{\pi^{2K}} \exp \left[-r^{T} \sigma r\right],
\end{align}
where $\sigma=(\Sigma+I_{4K})^{-1}$, and
$r=\left[X_1^A,P_1^A,...,X_K^B,P_K^B\right]^T$.
Noiseless attenuation alters $Q(r)$ to
\begin{equation}
\begin{aligned}
\frac{1}{P} e^{\left(T-1\right)\frac{[({X_1^A})^2+({P_1^A})^2]}{2}} Q([\sqrt{T}X_1^A,\sqrt{T}P_1^A,...,X_K^B,P_K^B]^T)\quad&\\
=\frac{\sqrt{\operatorname{det}(\tilde{\sigma})}}{\pi^{2K}} \exp \left[-r^{T} \tilde{\sigma} r\right],
\end{aligned}
\end{equation}
where $P$ is a normalization constant and
\begin{equation}\label{eq:cm_after_amp}
\tilde{\sigma}=
\begin{bmatrix}
T(\sigma_{1}-\frac{1}{2}I_{2})+\frac{1}{2}I_{2}& \sqrt{T}\sigma_{2}\\
\sqrt{T}\sigma_{3} & \sigma_{4}
\end{bmatrix},
\end{equation}
and where $\sigma_{1}, \sigma_{2}, \sigma_{3}, \sigma_{4}$ are sub-matrices of $\sigma$.
The CM of the state after noiseless attenuation can then be calculated by 
\begin{equation}
\tilde{\Sigma}=\tilde{\sigma}^{-1}-I_{4K}.
\end{equation}

Noiseless amplification alters the CM of a PDC state in a way similar to noiseless attenuation. 
To find the CM of the state after noiseless amplification one only needs to replace $\sqrt{T}$ in Eq.~(\ref{eq:cm_after_amp}) with $G$.
Additionally, for $\tilde{\Sigma}$ to be a valid CM the gain of the amplification operation must satisfy \cite{walk2013nondeterministic}
\begin{equation}\label{eq:limgain}
G<\sqrt{\frac{V+1}{V-1}},
\end{equation}
where $V$ is the variance of the quadratures of the supermode to be amplified.

\section{The Protocol for Multi-mode CV-QKD with noiseless attenuation followed by amplification}\label{sec:qkd}

\begin{figure}
	\centering
	\includegraphics[width=1\linewidth]{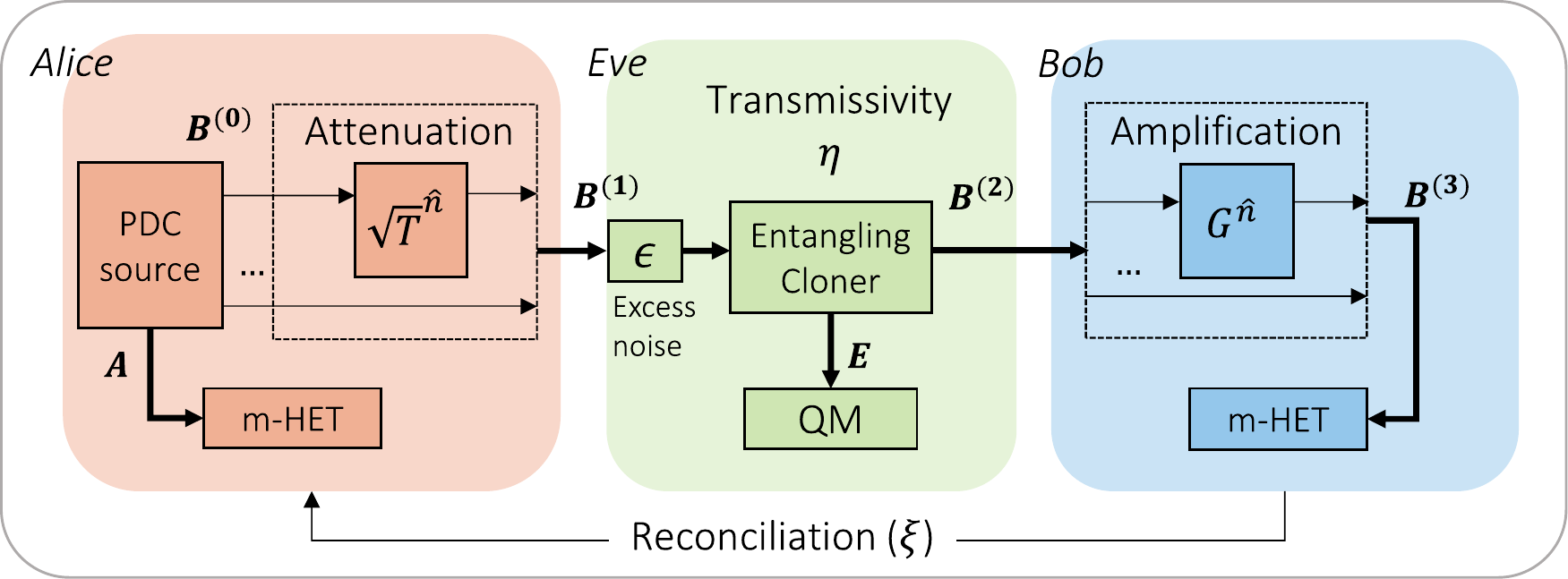}
	\caption{The system diagram of our multi-mode CV-QKD protocol. 
	A noiseless attenuation operation is applied to the first supermode of the PDC state at the transmitter (Alice), while a noiseless amplification operation is applied to the same supermode of the PDC state at the receiver (Bob).
	In the diagram, thick arrows represent the flow of the ensemble of supermodes (the entire beam) while thin arrows represent the flow of a single supermode.
	(m-HET: multi-mode heterodyne detection, QM: quantum memory.)}
	\label{fig:figdiagsystem}
\end{figure}

We build our multi-mode protocol upon an entanglement-based CV-QKD protocol with heterodyne measurements and reverse reconciliation \cite{weedbrook2004quantum}.
As illustrated in Fig.~\ref{fig:figdiagsystem}, Alice first prepares her PDC state ($ \bm {A}- \bm {B}^{(0)}$).
Again we assume this state only has $K$ equivalent EPR states.
Alice will apply a noiseless attenuation operation to the first supermode ($k=1$) of beam $\bm B^{(0)}$ while other supermodes are left unchanged.
The beam after attenuation, which is labeled as $ \bm B^{(1)}$, is sent to Bob via a satellite-to-ground channel controlled by Eve.

The channel is characterized by the transmissivity $\eta$ and the input excess noise $\epsilon$.
We assume $\eta$ is frequency independent and $\epsilon$ is i.i.d for each supermode.
Under our assumptions the supermode structure of $\bm B^{(1)}$ is retained after the channel. 
The multi-mode channel is equivalent to multiple independent sub-channels.
We assume Eve has full knowledge of Alice and Bob's protocol and has access to Bob's apparatus.
Eve will use the following strategy to steal the maximal available information.
For each sub-channel, Eve will first perform an {entangling cloner attack} \cite{weedbrook2012continuous} to obtain a purification of Alice and Bob's state.
She will then store her ensemble of purifications, $\bm E$, in her quantum memory.
Eve will perform a joint measurement on $\bm E$ after  reverse reconciliation.

The beam after the channel is labeled as $\bm B^{(2)}$.
Bob will perform noiseless amplification to the first supermode of $\bm B^{(2)}$.
The beam after amplification, labeled as $\bm B^{(3)}$, is then injected into a multi-mode heterodyne detector.

We assume a quantum memory device is available at Alice's side, so that she can prepare the PDC state with noiseless attenuation in advance. 
Under the assumption of infinite key size, the secret key rate for our multi-mode CV-QKD protocol is given by \cite{navascues2006optimality}
\begin{equation}\label{eq:keyratesub}
R_{\text{tot}} = \sum_{k=1}^{K}R_k,
\end{equation}
where
\begin{equation}
R_k = \xi I(A_k\negmedspace:\!B_k^{(3)}) - \chi(E_k\negmedspace:\!B_k^{(3)})
\end{equation}
is the secret key rate for each sub-channel, 
$\xi$ is the reverse reconciliation efficiency, $I(A_k\negmedspace:\!B_k^{(3)})$ is the classical mutual information between Alice and Bob, and $\chi(E_k\negmedspace:\!B_k^{(3)})$ is the Holevo bound for Eve's information.
These latter two quantities can be calculated using standard methods (see Appendix).

\section{The satellite-to-ground channels}\label{sec:channel}
\begin{figure}
	\includegraphics[width=.85\linewidth]{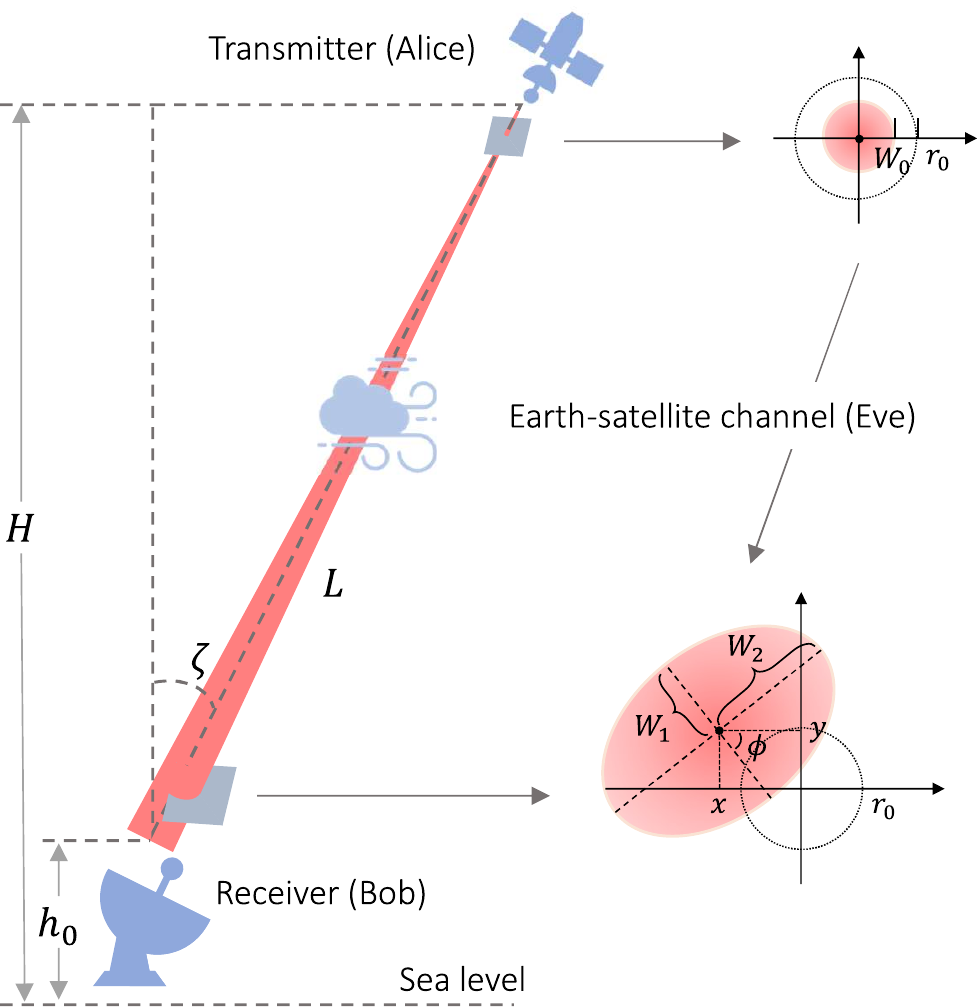}
	\centering
	\caption{System model for the satellite-to-ground channel.
		On the right-hand side of the figure, the red circle represents the beam-profile at the transmitter. The satellite-to-ground channel is assumed to be controlled by Eve, where beam-wandering, beam-broadening and beam-deformation alter the position and the shape of the beam-profile.
		The rotated red ellipse represents the beam-profile at the receiver.}
	\label{fig:channel_model}
\end{figure}

We consider the model of a down-link between a satellite  and a terrestrial station. 
The system model for the satellite-to-ground channel is depicted in Fig.~\ref{fig:channel_model}.
Our quantum information carrier is an ultra-fast pulsed optical beam.
First prepared at a satellite with altitude $H$ and zenith angle $\zeta$, the beam is sent to Bob through an atmospheric channel.
For optical signals in the atmospheric channel, the dominant loss mechanisms are beam-wandering, beam-broadening, and beam-deformation, all randomly caused by turbulence in the Earth's atmosphere \cite{andrews2005laser}. 
Beam-broadening is also a consequence of diffraction.
These effects are well-described by a model based on an elliptic-beam approximation \cite{vasylyev2016atmospheric}, which yields a reasonable agreement with experiments of short horizontal links under weak-to-moderate turbulence.
In this model, the beam intensity profile at the transmitter is characterized by five real random variables  $\left\lbrace x,y,\theta_1,\theta_2,\phi\right\rbrace$, where $(x,y)$ is the beam-centroid position, $\theta_i=\ln\frac{W_i^2}{W_0^2}$, $W_0$ is the beam-waist at the transmitter, $W_1$ and $W_2$ are elliptical semi-axis lengths, and $\phi$ is the rotation angle of the beam. 
The channel transmissivity reads
\begin{equation}
\eta = \eta_0\exp\left\lbrace-\left[\frac{\sqrt{x^2+y^2}/r_0}{R(\frac{2}{W_{\rm{eff}}(\phi-\phi_0)})}\right]^{ \lambda\left(2/W_{\rm{eff}}(\phi-\phi_0)\right) }\right\rbrace\textrm{,}
\label{Cheq1}
\end{equation}
where  $r_0$ is the aperture radius of the detector, $\phi_0=\tan^{-1}\frac{y}{x}$,  $W_{\rm{eff}}$ is the effective spot-radius, and $\eta_0$ is the maximal  transmissivity achieved when there is no beam-centroid deviation. 
These latter two parameters can be expressed by
\begin{equation}
\begin{aligned}
W_{\rm{eff}}^2(\phi)=  4r_0^2
\bigg\lbrace\mathcal{W}\Big(\frac{4r_0^2}{W_1W_2}&
e^{(r_0^2/W_1^2)\left[1+2\cos^2(\phi)\right]}\\ 
\times &e^{(r_0^2/W_2^2)\left[1+2\sin^2(\phi)\right]}\Big)\bigg\rbrace^{-1}\textrm{,}
\end{aligned}
\end{equation}
and
\begin{equation}
\begin{array}{*{30}{l}}
\eta_0 =
& 1 - I_0\left(r_0^2\left[\frac{1}{W_1^2}-\frac{1}{W_2^2}\right]\right)e^{-r_0^2\left(1/W_1^2+1/W_2^2\right)}\\
& - 2\left[1-e^{-(r_0^2/2)\left[1/W_1-1/W_2\right]^2}\right] \\
&\times\exp\left\lbrace-\left[\frac{\frac{(W_1+W_2)^2}{\left|W_1^2-W_2^2\right|}}{R\left(\frac{1}{W_1}-\frac{1}{W_2}\right)}\right]^{\lambda\left(\frac{1}{W_1}-\frac{1}{W_2}\right)}\right\rbrace.
\end{array}
\end{equation}
The scaling function $R(W)$ and the shaping function $\lambda(W)$ are given by,
\begin{equation}		
R(W) = \left[\ln\left(2\frac{1-\exp\left[-\frac{1}{2}r_0^2W^2\right]}{1-\exp\left[-r_0^2W^2\right]I_0(r_0^2W^2)}\right)\right]^{-\frac{1}{\lambda(W)}}\textrm{,}
\end{equation}	
and
\begin{equation}
\begin{aligned}
\lambda(W) =  2r_0^2&W^2 \frac{\exp{[-r_0^2W^2]}I_1(r_0^2W^2)}{1-\exp\left[-r_0^2W^2\right]I_0(r_0^2W^2)} \\ &
\times
\left[\ln\left(2\frac{1-\exp\left[-\frac{1}{2}r_0^2W^2\right]}{1-\exp\left[-r_0^2W^2\right]I_0(r_0^2W^2)}\right)\right]^{-1},
\end{aligned}
\label{Cheq5}
\end{equation}
respectively.
In the above equations, $\mathcal{W}(\cdot)$ is the Lambert $W$ function, and $I_i(\cdot)$ is the modified Bessel function of $i$-th order.

Building upon \cite{vasylyev2016atmospheric}, in \cite{guo2018channel} the elliptic-beam model is extended to satellite-to-ground links.
It is assumed that $x$ and $y$ are i.i.d. and they both follow a zero-mean Gaussian distribution. Parameters $\theta_1$ and $\theta_2$ are taken to follow a joint-Gaussian distribution. 
The rotation angle $\phi$ is uniformly distributed under the assumption that the turbulence is isotropic.
The mean and variance of $\left\lbrace x,y,\theta_1,\theta_2\right\rbrace$ are
\begin{equation}
\begin{aligned}
\left< \Delta x^2 \right>&=\left< \Delta y^2 \right>= 0.33W_0^2\sigma_I^2\Omega^{-7/6},\\
\left< \theta_{1} \right>&=\left< \theta_{2} \right>\\
&= \ln{\left[\frac{\left(1+2.96\sigma_I^2\Omega^{5/6}\right)^2}{\Omega^2\sqrt{\left(1+2.96\sigma_I^2\Omega^{5/6}\right)^2+1.2\sigma_I^2\Omega^{5/6}}} \right]},
\\
\left< {\Delta \theta_{1}}^2 \right>&=\left< {\Delta \theta_{2}}^2 \right>= \ln{\left[1+\frac{1.2\sigma_I^2\Omega^{5/6}}{\left(1+2.96\sigma_I^2\Omega^{5/6}\right)^{2}} \right]},
\\
\left< \Delta \theta_1 \Delta \theta_2 \right>&= \ln{\left[1-\frac{0.8\sigma_I^2\Omega^{5/6}}{\left(1+2.96\sigma_I^2\Omega^{5/6}\right)^{2}} \right]},
\\
\end{aligned}
\end{equation}
where  $\Omega=\frac{kW_0^2}{2L}$, $k$ is the optical wavenumber, $L$ is the propagation distance, and $\sigma_I^2$ is the scintillation index \cite{andrews2000scintillation}. 
The scintillation index can be written as
\begin{equation}
\sigma_I^2 = \exp{\left[ \frac{0.49\sigma_R^2}{\left(1+1.11 \sigma_R^{12/5}\right)^{7/6}} +  \frac{0.51\sigma_R^2}{\left(1+0.69\sigma_R^{12/5}\right)^{5/6}} \right]}-1\textrm{,}
\end{equation}
where $\sigma_R^2$ is the Rytov variance \cite{andrews2005laser},
\begin{equation}
\sigma_R^2 = 2.25k^{\frac{7}{6}}\sec^{\frac{11}{6}}{\zeta}\int_{h_0}^{H} C_n^2\left(h\right)\left( h-h_0 \right)^\frac{5}{6}\,dh
\label{sigmaR}
\end{equation}
with $h_0$ the altitude of the ground station and $C_n^2(h)$ the refraction index structure constant. This constant is described by the Hufnagel-Valley model \cite{beland1993propagation}
\begin{align}
C_n^2(h)=0.00594&(v/27)^2(h\times10^{-5})^{10}e^{-\frac{h}{1000}}\nonumber\\
&+2.7\times 10^{-16}e^{-\frac{h}{1500}}+Ae^{-\frac{h}{100}}\textrm{,}
\label{lasteq}
\end{align}
where $v$ is the r.m.s. wind speed in ${m/s}$ and $A$ is the nominal value of $C_n^2(0)$ at sea level in ${m}^{-2/3}$.

\section{Simulation}\label{sec:results}
\subsection{Simulation Settings}
For each sub-channel, unless otherwise specified, we set the channel input excess noise $\epsilon=0.05$ (in vacuum noise unit) and the reconciliation efficiency $\xi=0.95$.
We assume the number of equivalent EPR states of the PDC state created by Alice is $K=5$.
These EPR states are characterized by the squeezing parameters $\left[r_1,r_2,...,r_5\right]=g\left[\lambda_1,\lambda_2,...,\lambda_5\right]$ defined in Eq.~(\ref{eq:EPRsingle}).
For the normalized coefficients of the PDC state $\lambda_1,\lambda_2,...,\lambda_5$, we consider three scenarios.
In the first scenario the $\lambda_k$'s are all zero except $\lambda_1$. This state is a good approximation to a single-mode state.
In the second scenario the $\lambda_k$'s follow an exponentially decaying distribution. 
We refer to this as a \emph{generic} supermode system.
In the last scenario the $\lambda_k$'s are all identical.

For the satellite-to-ground channel, we adopt the parameters from \cite{guo2018channel}. These are $h_0=0$, $v=6\text{m/s}$, and $C_n^2(0)=9.6\times10^{-14}\text{m}^{-2/3}$.
The beam waist at the transmitter and the receiver aperture are set to $W_0=6\text{cm}$ and $r_0=1\text{m}$, respectively.
{Under these settings the mechanism that dominates the channel loss is beam-broadening.}
In accordance with the experiments in \cite{roslund2014wavelength}, the center wavelength of the multi-mode beam is set as $\lambda=795\text{nm}$ with a 30dB bandwidth of approximately $20\text{nm}$ (6nm of FWHM). The pulse rate of the beam is 76MHz.
We assume that all the frequency components of the beam undergo the same attenuation as the central frequency component.

For the practical implementation of our CV-QKD protocol we need to consider the success probabilities for noiseless attenuation and noiseless amplification. 
The success probability for noiseless attenuation can be viewed as unity since we assume Alice can prepare the attenuated PDC state in advance and then store the state in quantum memory.

{
In practice noiseless amplification is impossible, 
but there are various procedures that can approximate noiseless amplification (e.g., \cite{mivcuda2012noiseless,chrzanowski2014measurement,ulanov2015undoing,zhao2017characterization,winnel2020generalised}).
One of these procedures assumes noiseless amplification is only performed on the subspace of the first $N+1$ Fock states \cite{mivcuda2012noiseless}. 
The success probability for this procedure is lower bounded by
$G^{-2N}$.
In the simulation to follow, we will set the success probability for noiseless amplification by $P_k= G^{-2\lceil\bar{n}\rceil}$, where $\bar{n}$ is the average photon number of the supermode to be amplified at the receiver.
The total key rate defined in Eq.~(\ref{eq:keyratesub}) is then re-written as\footnote{For the PDC states we considered the average photon number of the supermode to be amplified at the receiver satisfies $\bar{n}<1$.
For example, when $\eta=10^{-3}$ (30dB) we have $\bar{n}\sim2\times10^{-3}$.
We will ignore the non-Gaussianity induced by the procedure that approximates noiseless amplification. 
At low average photon numbers ($\bar{n}\ll1$) it is easy to show that this procedure leads to a negligible impact on key rates.}
\begin{equation}\label{eq:keyratepk}
\begin{aligned}
R_{\text{tot}} = \sum_{k=1}^K P_k\left[\xi I(A_k\negmedspace:\!B_k^{(3)}) - \chi(E_k\negmedspace:\!B_k^{(3)})\right].
\end{aligned}
\end{equation}
}


\subsection{Simulation Results}
\begin{figure}
	\centering
	\includegraphics[width=0.92\linewidth]{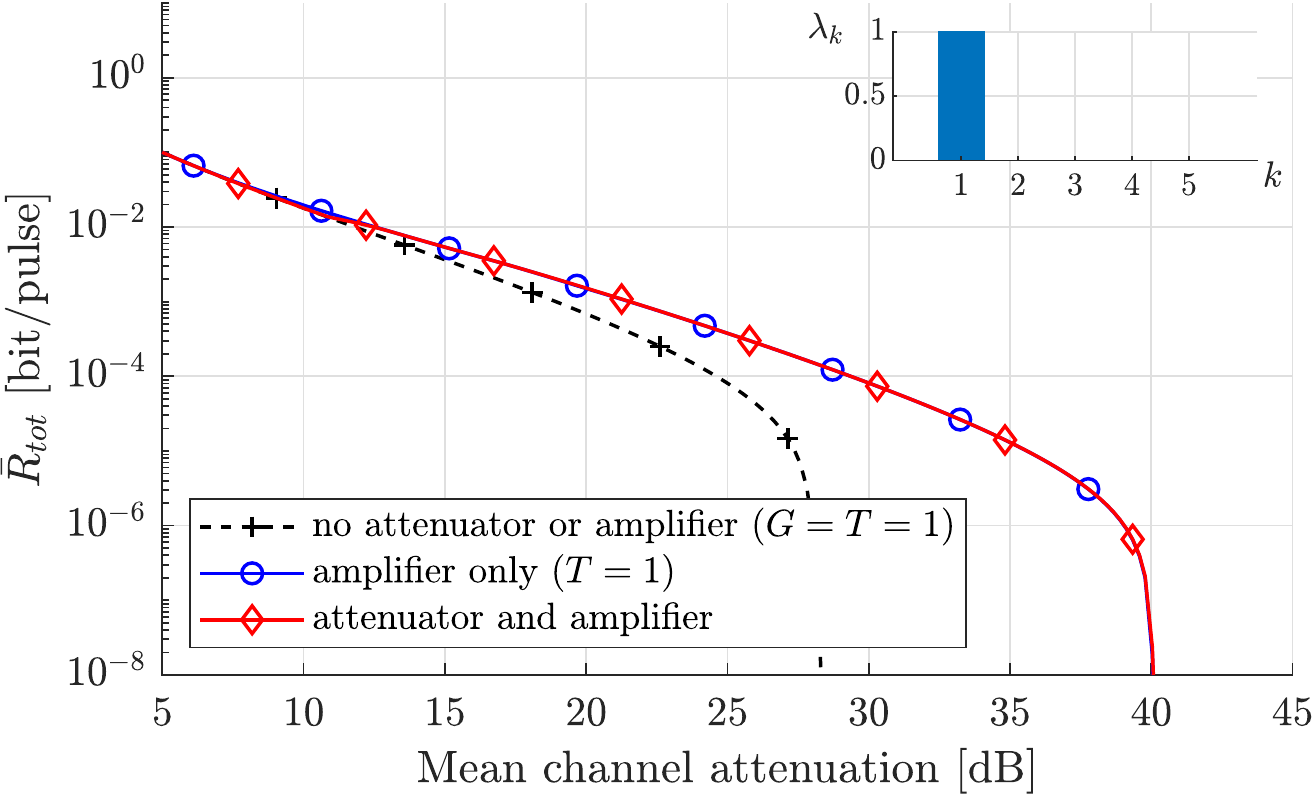}
	\includegraphics[width=0.92\linewidth]{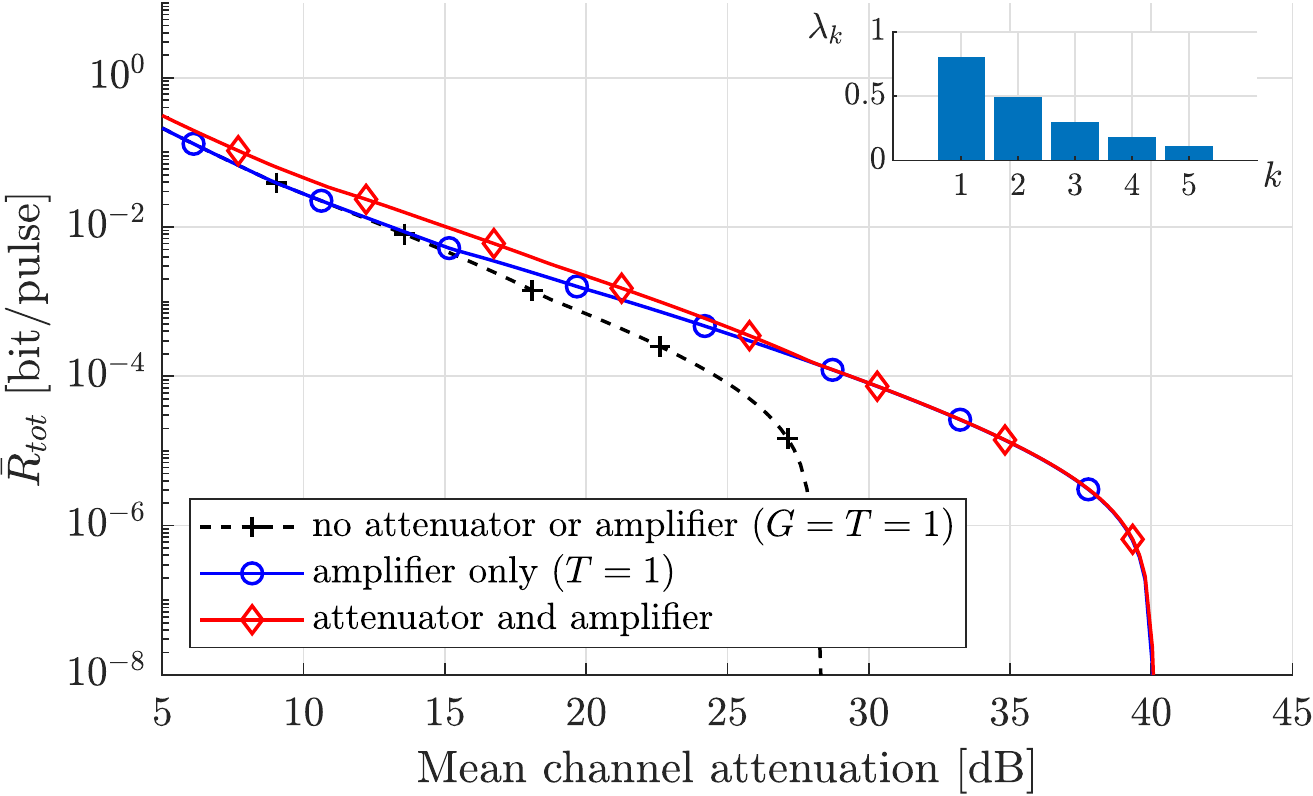}
	\includegraphics[width=0.92\linewidth]{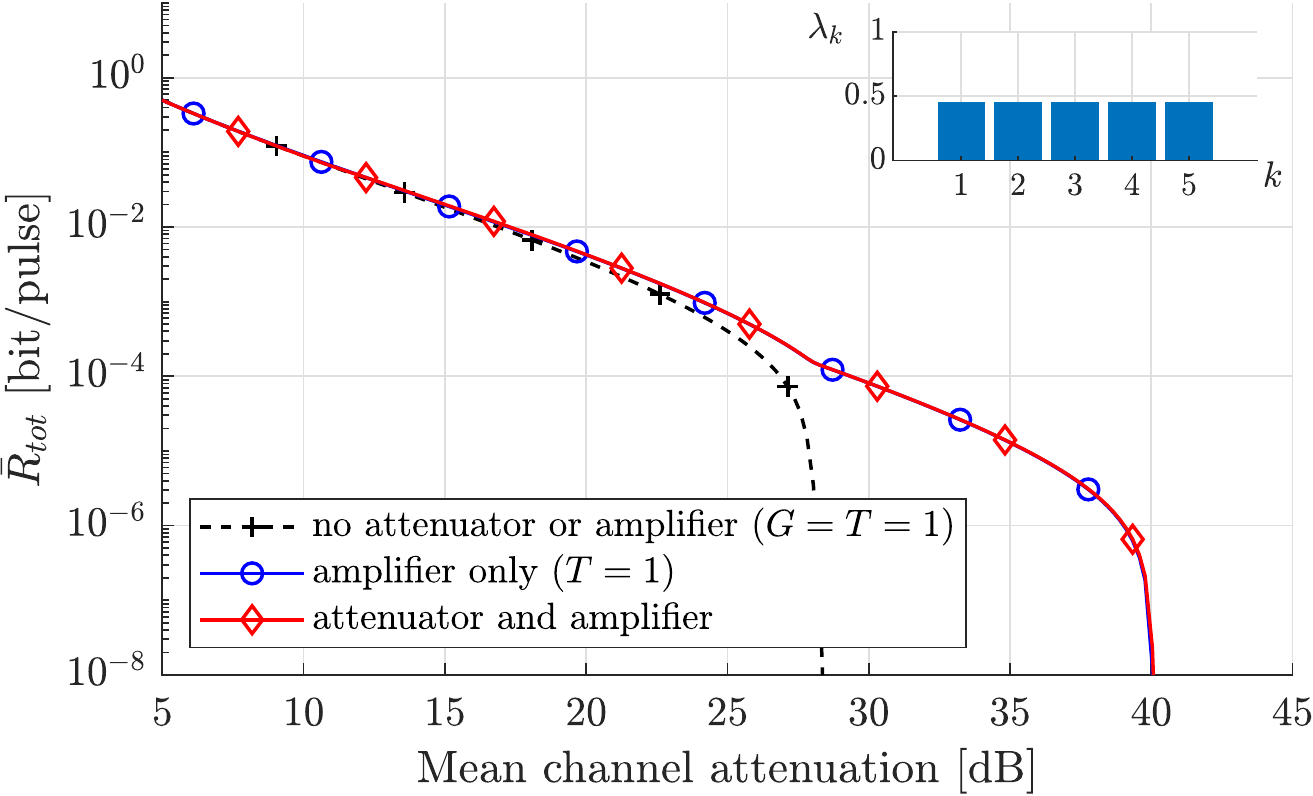}
	\caption{Maximized key rate over satellite-to-ground channels, where noiseless attenuation and amplification are both applied to the first supermode. 
	The black dashed curve represents the key rate with no attenuator or amplifier, the blue solid curve represents the key rate with only an amplifier, and the red solid curve represents the key rate with an attenuator and an amplifier. 
	The inset illustrates the supermode structure of the initial PDC state.}
	\label{fig:figattamp11}
\end{figure}

\begin{figure}
	\centering
	\includegraphics[width=0.95\linewidth]{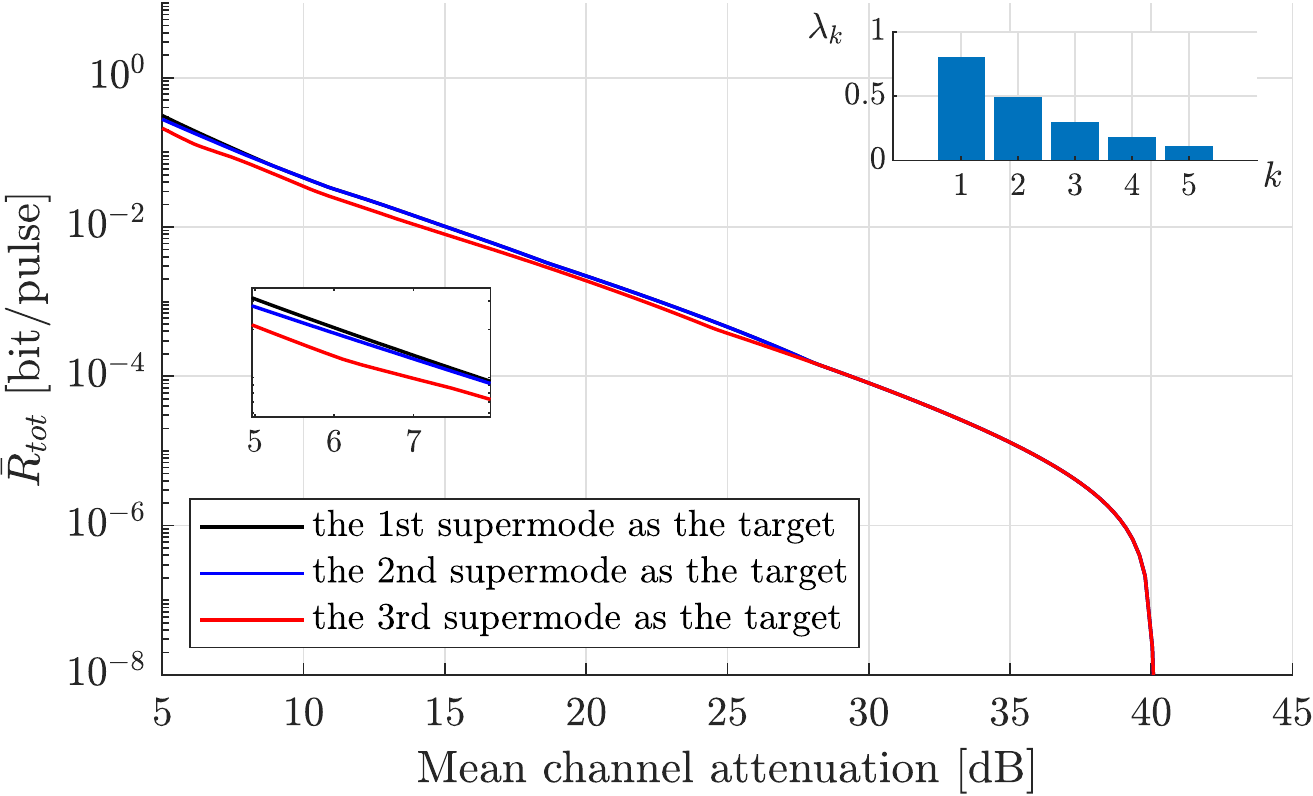}
	\caption{Maximized secret key rate against mean channel attenuation, where noiseless amplification and attenuation are applied to the first supermode (black), the second supermode (blue), or the third supermode (red). The top-right inset illustrates the supermode structure of the PDC state.}
	\label{fig:figattampother}
\end{figure}

We first investigate the strategy where Alice performs noiseless attenuation to one supermode, while Bob performs noiseless amplification to the same supermode. We refer to this strategy as the \textit{symmetrical} strategy.
For each channel attenuation level, we maximize the total key rate defined in Eq.~(\ref{eq:keyratepk}) on the PDC gain of the original state, $g$, the transmissivity of the noiseless attenuator, $T$, and the gain of the noiseless amplifier, $G$.
For satellite-to-ground channels we use the averaged secret key rate, $\bar R_{\text{tot}}$, as our performance metric. Since the PDF for channel transmissivity is in general intractable, we calculate $\bar R_{\text{tot}}$ by
$\bar R_{\text{tot}}=\frac{1}{N_{\text{sample}}}\sum_n R_{\text{tot}}(\eta_n),$
where $\eta_n$ is the channel transmissivity sample generated by a Monte Carlo algorithm and $N_{\text{sample}}$ is the number of samples.
We assume the channel transmissivity is measured within each coherence time window and $\left\lbrace g,G,T \right\rbrace$ are optimized based on this measurement.
The results are illustrated in Fig.~\ref{fig:figattamp11} and Fig.~\ref{fig:figattampother}, where the mean channel attenuation is calculated by
\begin{equation}\label{eq:meanatt}
\bar \eta [\text{dB}]=-10 \log_{10} \left[\frac{1}{N_{\text{sample}}}\sum_n (\eta_n)\right].
\end{equation}

In Fig.~\ref{fig:figattamp11} we compare the maximized $\bar R_{\text{tot}}$ for the multi-mode CV-QKD protocol with or without amplification and attenuation applied to the first supermode.
For the most likely PDC sources (the middle figure in Fig.~\ref{fig:figattamp11}), applying noiseless amplification to the first supermode can significantly increase the maximized $\bar R_{\text{tot}}$ when the mean channel attenuation is large (>28dB for the channel parameters we considered).
Such improvement can be further enhanced by adding a noiseless attenuator at the transmitter. 
We note that the improvement offered by noiseless attenuation cannot be observed in single-mode cases.
In practice, noiseless attenuation is not needed for single-mode states since it amounts to reducing the squeezing of the original state, which can be easily realized by directly adjusting the PDC gain.
In Fig.~\ref{fig:figattampother} we compare the maximized $\bar R_{\text{tot}}$ for the situation where noiseless amplification and attenuation are both applied to a supermode other than the first supermode. 
Results show that applying the operations to the first supermode offers the largest $\bar R_{\text{tot}}$ over the entire range of mean channel attenuation we have considered.

We then study the strategy where Alice performs  noiseless attenuation to the first supermode, while Bob performs noiseless amplification to a different supermode. The results are almost identical to the results of the symmetrical strategy ($<5\%$ inferior).

\begin{figure}
	\centering
	\includegraphics[width=0.45\linewidth]{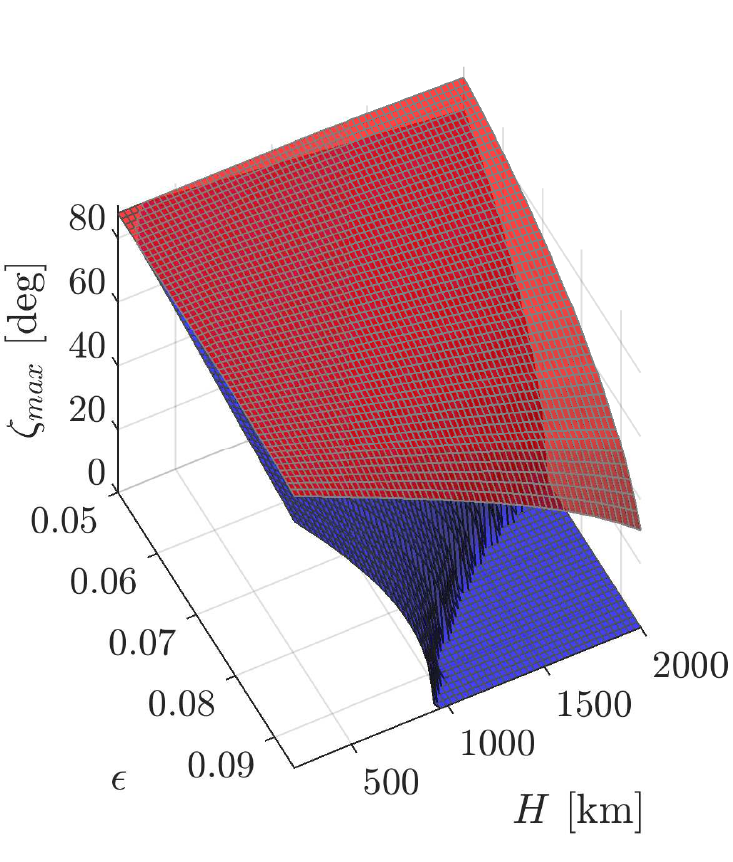}
	\includegraphics[width=0.45\linewidth]{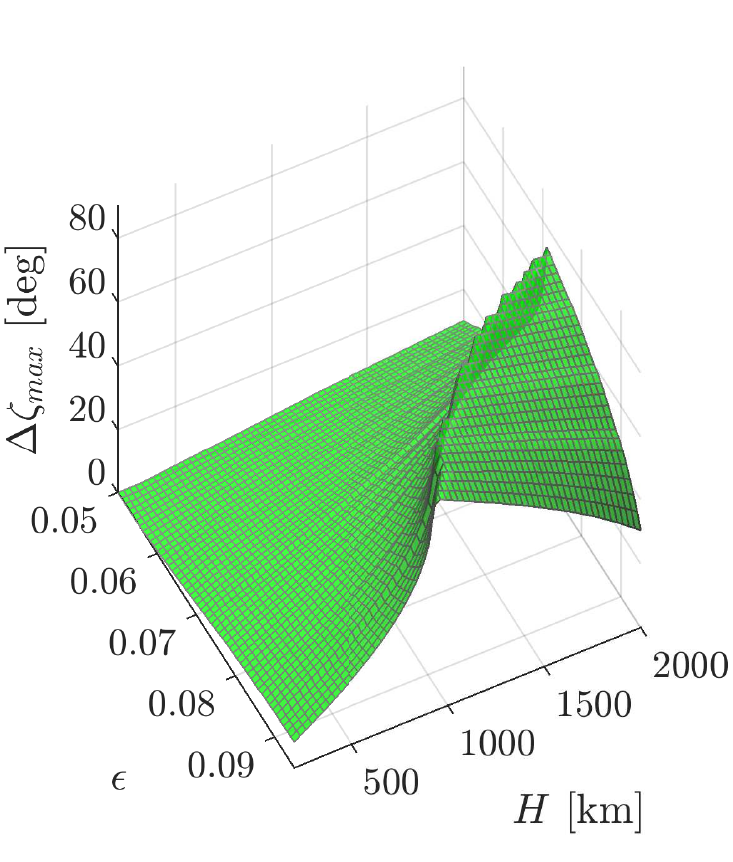}
	\\
	\includegraphics[width=0.45\linewidth]{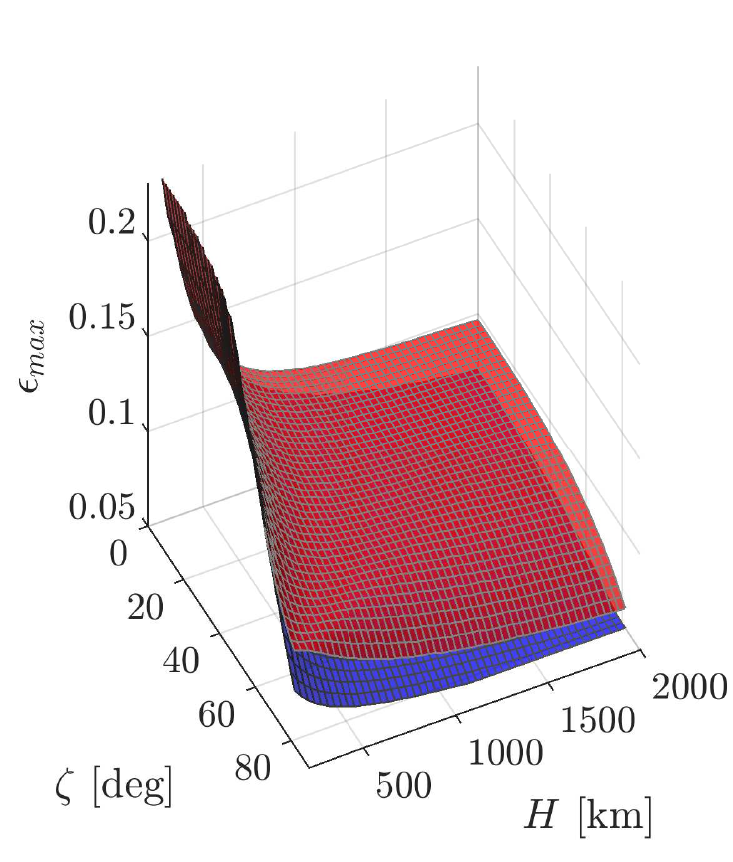}
	\includegraphics[width=0.45\linewidth]{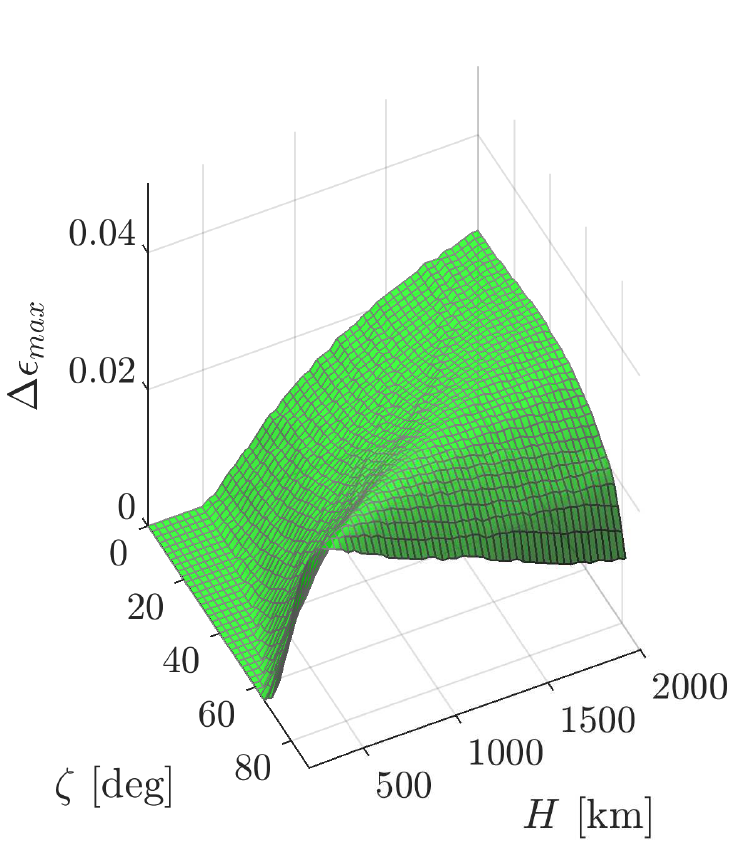}
	\caption{Top panel: Maximal acceptable zenith angle against the altitude of the satellite and the channel input excess noise.
		Bottom panel: Maximal tolerable channel input excess noise against the altitude of the satellite and the zenith angle.
		(Red surface: the improved protocol with attenuation and amplification. 
		Blue surface: the original protocol without attenuation or amplification. 
		Green surface: the improvement offered by the improved protocol.)}
	\label{fig:maxzenithangle}
\end{figure}

We also investigate the maximal acceptable zenith angle $\zeta_{\text{max}}$ and the maximal tolerable channel input excess noise $\epsilon_{\text{max}}$ that allow for a positive  $\bar R_{\text{tot}}$.
Focusing on the symmetrical strategy with generic PDC sources, in the top figure of Fig.~\ref{fig:maxzenithangle} we compare $\zeta_{\text{max}}$ for protocols with or without amplification and attenuation against the altitude of the satellite $H$ and the channel input excess noise $\epsilon$.
Results show that our improved protocol (with amplification and attenuation) offers an increased $\zeta_{\text{max}}$ over the entire range of the altitude of an LEO satellite ($<2000$km). Such an improvement is more significant when  $\epsilon$ is large.
In the bottom figure of Fig.~\ref{fig:maxzenithangle} we compare $\epsilon_{\text{max}}$ against the satellite altitude and the satellite zenith angle.
It can be seen that our improved protocol can significantly enhance $\epsilon_{\text{max}}$ when the channel condition is bad (i.e., large satellite altitude and zenith angle). 
The enhancement is insignificant when the mean channel attenuation is small due to the limit on the maximal gain for noiseless amplification (Eq.~(\ref{eq:limgain})). 

{ We note that the results provided here for satellite-to-ground channels can be directly related to fixed-attenuation channels such as optical fibers. 
This is achieved through the relation from which the mean channel attenuation is determined (Eq.~(\ref{eq:meanatt})).
For example, in cases where the PDF of the satellite-to-ground channel approaches a delta function, the results of $\bar R_{\text{tot}}$ for mean attenuation approach those of fixed attenuation. 
The PDF of the satellite-to-ground channel is heavily depended not only on the turbulence parameters but also on the dimension of the receiver aperture. 
In many instances, e.g. when the received beam dimension is much larger than the receiver aperture, the mean attenuation results will be very close to the fixed attenuation results.}

\section{Conclusion}\label{sec:conclusion}
In this work, we investigate the use of noiseless attenuation and noiseless amplification, in the context of entanglement-based multi-mode CV-QKD over satellite-to-ground channels.
We find that noiseless amplification can significantly increase the noise tolerance and the transmission distance, allowing for satellites with higher altitudes and hence extended coverage.
Interestingly, while noiseless attenuation is replaceable with a less-squeezed EPR state in the single-mode setting, in the multi-mode setting, for generic multi-mode entangled sources noiseless attenuation is irreplaceable.
Our results will be particularly important in future space-based missions, in which quantum memory forms part of the satellite payload and broadband pulses of lights are utilized.

\section*{Acknowledgment}
Mingjian He is partially supported by a Postgraduate award through the China Scholarship Council.

\section*{Appendix}
For each sub-channel (indexed by $k$), Alice and Bob's mutual information can be calculated by
\begin{equation}
I(A_k\negmedspace:\!B_k^{(3)})=\log_2{\frac{V_{A_k}}{V_{A_k|B_k^{(3)}}}}\textrm{,}
\label{eq:abmutual}
\end{equation}
where $V_{A_k}$ is the variance of Alice's supermode, and ${V_{A_k|B_k^{(3)}}}$ is the variance of the quadratures of Alice's supermode conditioned on Bob's heterodyne measurement.
The Holevo bound for Eve's information is
\begin{equation}\label{eq:eveinfo}
\chi(E_k\negmedspace:\!B_k^{(3)}) = g\left( \alpha_{1,k}\right) + \left( \alpha_{2,k}\right) - g\left( \alpha_{3,k}\right)\textrm{,}
\end{equation}
where
$g(x)=\frac{x+1}{2}\log_2\frac{x+1}{2}-\frac{x-1}{2}\log_2\frac{x-1}{2}$, $\alpha_{1,k}$ and $\alpha_{2,k}$ are the symplectic eigenvalues of the CM of state $\rho_{AB^{(3)}}$, and $\alpha_{3,k}$ is the symplectic eigenvalue of the CM of Alice's supermode conditioned on Bob's measurement.

The quantities in Eqs.~(\ref{eq:abmutual}) and (\ref{eq:eveinfo}) are fully determined by the CM of Alice and Bob's state at different stages.
Noticing the evolution of the CM follows the same procedure for each sub-channel, we use the first EPR state ($k=1$) as an example to derive the CMs.
Let
\begin{align}
\Sigma_{A_1B_1^{(1)}}=\left(\begin{array}{cc}
{a I_2} & {c Z} \\
{c Z} & {b I_2}\end{array}\right),
\end{align}
be the CM of the EPR state after the noiseless attenuation at the transmitter (the exact form of $\Sigma_{A_1B_1^{(1)}}$ can be calculated using Eq.~(\ref{eq:cm_after_amp}), which we omit here for conciseness).
This state is stored in Alice's quantum memory before being sent to Bob.
The channel alters the above CM to
\begin{equation}\label{eq:cmeig1}
\Sigma_{A_1B_1^{(2)}} = \left[ {\begin{array}{*{20}{cc}}
	aI_2 &\sqrt{\eta}cZ \\
	\sqrt{\eta}cZ&\left[\eta(b+\epsilon)+(1-\eta)\right] I_2  \\
	\end{array}}\right],\\
\end{equation}
Bob will perform the noiseless amplification on his received state. Let
\begin{align}
\Sigma_{A_1B_1^{(3)}}=\left(\begin{array}{cc}
{x I_2} & {z Z} \\
{z Z} & {y I_2}\end{array}\right),
\end{align}
be the CM of Alice and Bob's state $\rho_{AB^{(3)}}$ after the noiseless amplification (the exact form of $\Sigma_{A_1B_1^{(3)}}$ can also be calculated using Eq.~(\ref{eq:cm_after_amp})).
The symplectic eigenvalues of the above CM, $\alpha_{1,1}$ and $\alpha_{2,1}$, can be calculated by
\begin{equation}
\begin{aligned}
\alpha_{1,1}=\frac{1}{2}\left[ \sqrt{(x+y)^2-4z^2}+(y-x)\right],\\
\alpha_{2,1}=\frac{1}{2}\left[ \sqrt{(x+y)^2-4z^2}-(y-x)\right].
\end{aligned}
\end{equation}

The CM of Alice's supermode conditioned on Bob's heterodyne measurement is
\begin{equation}\label{eq:cmeig2}
\Sigma_{A_1|B_1^{(3)}}=
\left(x-\frac{z^2}{y+1}\right) I_2,
\end{equation}
which has the symplectic eigenvalue $\alpha_{3,1}=x-\frac{z^2}{y+1}$.
The mutual information
can now be expressed as
\begin{equation}
\begin{aligned}
I(A_k\negmedspace:\!B_k^{(3)})
=\log_2{\frac{(x+1)(y+1)}{(x+1)(y+1)-z^2}}.
\end{aligned}
\end{equation}
The Holevo bound for Eve's information can be calculated by putting $\alpha_{1,1}$, $\alpha_{2,1}$, and $\alpha_{3,1}$ into Eq.~(\ref{eq:eveinfo}).


\bibliographystyle{IEEEtran}

\end{document}